\documentclass[twocolumn,aps, prd, floats,floatfix, superscriptaddress, nofootinbib, showkeys]{aastex631}

\usepackage{amsmath} 
\usepackage{color,soul}
\usepackage{float}

\begin{document}

\title{Luminosity Function of Galaxies in Voids: A Modification Inspired by Excursion Set Theory}

\author{Zeynab Ashurikisomi}
\affiliation{Department of Physics, Sharif University of Technology, Tehran 11155-9161, Iran}
\author[0000-0001-6131-4167]{Shant Baghram}
\email{baghram@sharif.edu}
\affiliation{Department of Physics, Sharif University of Technology, Tehran 11155-9161, Iran}
\affiliation{Research Center for High Energy Physics, Department of Physics, Sharif University of Technology, P.O.Box 11155-9161, Tehran, Iran}
\date{\today}

\begin{abstract}
In the standard picture of cosmology, the galaxies reside in dark matter (DM) halos. DM halos are distributed in the cosmic web in different environments. The luminosity of the galaxies in different environments can be used as a probe to assess a cosmological model. This study focuses on the properties of galaxies in void regions, where halos typically do not experience extreme conditions. By examining the galaxy luminosity function, we aim to understand the dependence of galaxy properties on their environment and redshift so that later, we can use this as a tool to evaluate cosmological models. We employ the excursion set theory to incorporate parameters related to the number density of DM halos into the luminosity function. 
Using the Galaxy and Mass Assembly (GAMA) survey and 2dFGRS  datasets, we fit our theoretical models to observational data, examining the environmental and redshift dependence of the galaxy luminosity function. Our results indicate that we model the galaxy luminosity function in voids effectively by considering the linear density contrast of the environment and the growth function $D(z)$ for redshift dependence. This study provides a model for the environmental dependence of galaxy luminosity function that offers an improvement in the $\chi ^2$ parameter compared to the previously proposed model in \cite{mcnaught2014galaxy}. Both Bayesian information criterion (BIC) and Akaike information criterion (AIC) tests support the superiority of this model for the void region. 
\end{abstract}



\section{Introduction} \label{sec:intro}
In the standard picture of structure formation, galaxies reside in the dark matter (DM) halos \citep{white1978core,Wechsler_2018,Green_2022}. A generic consequence of this could be that the properties of the galaxy population are closely related to those of DM halos \citep{Scholz_D_az_2022,Zhang_2022}.
The dependence of the local processes {on smaller scales} like active galactic nuclei (AGN), supernovae explosions, and stellar winds on galaxy evolution has been widely studied \citep{roos1981galaxy,morganti2017,fontanot2020}.
Aside  from  these  processes,  understanding  the  dependence of the properties of the galaxies on that of the DM halos is of great importance \citep{balletiii2021}.
It is interesting to note that the environment in which a halo exists can also be tightly related to the properties of the galaxy it hosts. So different galaxy properties, such as color, size, mass function, metallicity, and luminosity function, could be different for a void galaxy from the galaxy residing in an overdense region \citep{curtis2024}. 

Among the different {properties of galaxies}, the luminosity function, $\phi(L)$, is one of the most accessible observables to measure the distribution of galaxies in the Universe \citep{croton20052df}. The recent observations and cosmological simulations demonstrate the ongoing efforts to understand the galaxy luminosity function and its role in shaping our knowledge of galaxy formation and evolution. One recent study presented the rest-frame ultraviolet galaxy luminosity function and luminosity density measurements in the far-UV wavelength, in the redshift range of $0.6 - 1.2$, using deep UV imaging of the Chandra Deep Field South \citep{sharma2022}. \cite{sabti2022} focused on the high-redshift galaxy luminosity function and its role in studying the physics at play during the epoch of reionization.

The dependence of galaxy luminosity function on different factors such as color, redshift, and environment has been widely investigated \citep{loveday2012galaxy,mcnaught2014galaxy, bouwens2022z}. For example, \cite{tavasoli2015} have investigated the relationship between the galaxy luminosity function and the overdensity of the environment and galaxy color in underdense regions. 

In this work, we specifically focus on the properties of galaxies residing in void regions. The halos residing in voids usually do not experience extreme conditions such as ram pressure stripping, accretion, and merger \citep{tavasoli2015}. So, in these environments, we can focus on the effect of cosmology and environment on structure formation.

To link the galaxy properties to the halos in which they reside, we need a procedure for halo occupation distribution (HOD). HOD is expressed as a conditional probability $P(N|M_h)$, which is the probability of finding $N$ galaxies with some specified properties in a DM halo of mass $M_h$.
Numerous investigations illustrate that the HOD insights are exceptionally capable of portraying the association between galaxies and DM halos \citep{yang2009galaxy}.

In this work, we try to investigate a theory-based analysis of the dependence of galaxy luminosity function on the redshift and environment. 
For this task, we were inspired by a nonlinear structure formation model called excursion set theory (EST) \citep{bond1991,Sheth:1999su,Zentner:2006vw,Nikakhtar:2016bju} {to model environment (or redshift)-dependence of the luminosity function. These parameters are related to the number density of DM halos in different environments (redshifts).}  
This manuscript is organized as follows. In Section \ref{Sec2}, we review the relations linking galaxy luminosity function to halo abundance. In Section \ref{Sec3}, we discuss the environment and redshift dependence of galaxy luminosity function, where we present our results, and finally, in Section \ref{Sec4}, we conclude.


\section{theoretical background} \label{Sec2}
In this section, we will review the physics of DM halo distribution and the occupation of halos by galaxies. In the first two subsections, we study halos and voids in nonlinear structure formation, and in the third one, we will review the idea of linking galaxies with halos.
\subsection{ DM Halos and Voids } \label{Sec2.1}
To find the statistics of DM halo structure, we need to deviate from the linear regime, in which the amplitude of the matter density perturbation is assumed to be small. \cite{press1974formation} introduced an insightful idea to find the abundance of these overdense halo regions, which later became the basis of many more realistic and complex halo formation models. In their model, they assumed that any spherical comoving region in which the smoothed linear overdensity exceeds a critical value $\delta_c \simeq 1.69$ will be considered as a DM halo. This assumption was later logically enhanced by \cite{bond1991} in a more sophisticated nonlinear structure formation model such as Excursion Set Theory (EST). In the context of EST, this framework leads to the following relation that describes the number density of the DM halos at redshift $z$  with mass greater than $M$.
\begin{eqnarray}
    \label{eq:1}
	n(M,z) &=& \int_M^{\infty}{\sqrt{2\over{\pi}}}{{\rho_{\text{\tiny{M}}}}\over{M^2}}\nu(M,z) {{d \ln{\nu(M,z)}}\over{d \ln{M}}}  \\ \nonumber
	&\times &\exp[-{{{\nu^2(M,z)}}\over{2}}\big]dM,
\end{eqnarray}
where $\nu(M,z) = \delta_c / \sigma(M,z)$ is the height parameter and  $\sigma(M,z)$ is the mass variance {(on scale R corresponding to mass M using a desired window function \textemdash  in EST, the sharp-k window function is used)} encapsulating the cosmological model used in theory \citep{Zentner:2006vw,Kameli:2020kao,Kameli:2025qyq}. Both the scale and redshift dependence of the mass function emerge from the height parameter.
It has to be mentioned that the aforementioned critical value is a linearly extrapolated overdensity. Since the EST is written in the Lagrangian framework, the physical (Eulerian) equivalent value can be calculated using mass conservation.\cite{mo1996} have extracted the following relation that can be used to transform the Lagrangian density contrast $\delta_L$ to the Eulerian one $\delta_{\text{NL}}$ as below,
\begin{eqnarray}
    \label{eq:2}
	\delta_L(\delta)&=&1.68647-\frac{1.35}{(1+\delta_{\text{NL}})^{2 / 3}}-\frac{1.12431}{(1+\delta_{\text{NL}})^{1 / 2}}  \\ \nonumber &+&\frac{0.78785}{(1+\delta_{\text{NL}})^{0.58661}}.
\end{eqnarray}

As discussed, the cosmic web has different components, such as nodes (halos), filaments, sheets, and voids, and among them, the void region recently attracted a lot of attention \citep{libeskind2018tracing,hossen2022ringing}. That is why we need a similar approach to obtain the number density of voids in the context of the EST.
To propose a theoretical model for voids, in the context of EST, we can once more assume that our void has a spherical shape and the linear underdensity denoted to the formation of such regions is approximately $\delta_v \simeq -2.81$ as imposed by the shell crossing constraint \cite{sheth2004hierarchy}. The pivotal point here is that one needs to consider that there might be a void forming inside a larger halo. The collapse of such a halo will destroy such voids, which is when the concept of two barriers gets highlighted in the EST. Using these two barriers, one can ensure that no such voids are considered in the void number density. We define a parameter to include the height difference of these two barriers $   \mathcal{D} \equiv {|\delta_v| \over \delta_c +|\delta_v|}$. 
With similar calculations, \cite{sheth2004hierarchy} have found the void distribution:
\begin{eqnarray}
	\label{eq:3}
	SF_v(S, \delta _c, \delta _v)&=&\sum_{j=1}^\infty {j^2 \pi^2 \mathcal{D}^2\over{\delta_v^2/S}}{\sin(j\pi \mathcal{D} )\over {j \pi}} \\ \nonumber
	&\times&\exp(-{j^2 \pi^2 \mathcal{D}^2\over{2\delta_v^2/S}}),
\end{eqnarray}
{in which $F_{v}$ is the distribution of the first crossings of the random walks associated with halos in voids and S is the mass variance, $S\equiv \sigma ^2$. In equation} \ref{eq:3} {$F_v$ determines the fraction of mass inside the voids that contain mass $M(S)$. Again, the relation between mass and variance comes from the applied window function. For convenience, this interdependence between mass($M$) and scale ($S$ or $R$) has not been shown in equation} \ref{eq:3}. {Then, the number density of voids with radius larger than $R$ (or, equivalently with mass higher than $M$) can be obtained. }
\begin{eqnarray}
	n_{\rm{void}}(R, z)&=&  \int_R^{\infty} {dn_{\rm{void}}\over{dR}}dR \\ \nonumber
	&=& \int_R^{\infty} {{\bar{\rho}_v} \over{M(R)}}|{dF_v(S(R), \delta _c, \delta _v)\over{dR}}|dR.
	\label{eq:4}
\end{eqnarray}
In this equation, $\bar{\rho}_v$ is the average DM density inside voids. In the next subsection, we will study the physics of halos in voids.

\subsection{Halo in Void}\label{Sec2.2}
The EST can be extended to give us the number density of halos residing inside a radially larger void \citep{Kameli:2025qyq}. For convenience, we call this type of DM halo "halo in void". The idea is that we will again consider two barriers, one corresponding to halo formation $\delta_c$, and one corresponding to void formation criteria $\delta_v$. But this time, we demand a condition in which the sphere corresponding to the host void is larger than the sphere corresponding to the DM halo. Using the same approach as the EST, it has been shown that the number density of halo in void $ n _{\rm{hv}}$ can be written as follows \citep{parkavousi2023voids}.
\begin{eqnarray}
	n _{\rm{hv}} &=& \int{\bar{\rho_v}\over{M\sqrt{2\pi}}}{\delta_c(z)-\delta _v(z) \over{({S_{\rm{h}}(M)-S_v (R)})^{3/2}}} \\ \nonumber &\times& \exp\left[-{(\delta_c(z)-\delta_v (z))^2 \over{2(S_{\rm{h}}(M)-S_v(R))}}\right]{dS\over{dM}}dM,
	\label{eq:5}
\end{eqnarray}
in which we have defined $S\equiv\sigma^2(R)$, where the indices $\rm{h}$ and $v$ are denoted to the variance of the scales corresponding to halos and voids. This relation is a consequence of the conditional probability of two barrier crossings. The first barrier is related to the void $\delta_v$ and the second to the halo $\delta_c$.


\subsection{Linking galaxy luminosity function to DM halos}
\label{Sec2.3}
The galaxy luminosity function is usually modeled by the Schechter function, constituting a power law multiplied by an exponential cut-off at high luminosity \citep{schechter1976analytic}. The function has the following form:
\begin{equation} \phi\left(L\right) dL=\phi_*{\big(\frac{L}{L_*}\big)}^\alpha \exp[{(-L/L_*)}] d(\frac{L}{L_*}),
	\label{eq:6}
\end{equation}
in which $L_*$, $\phi_*$, and $\alpha$ are free parameters. On the other hand, the conditional luminosity function, 
$\phi(L|M_{\rm{h}},z)dL$, which is an instance of HOD, is defined as the average number of galaxies with luminosity between $L$ and $L+dL$ residing inside halos with mass $M_{\rm{h}}$ in a specific redshift $z$. So one can extract the galaxy {luminosity function} through the following relation \citep{mo2004dependence}:

\begin{equation}
	\Phi (L, z) = \int \Phi(L|M)\;n(M,z)dM,
	\label{eq:7}
\end{equation}
where $n(M,z)$ is the mass function of the DM halos at redshift z, for which there exist several analytical models {pioneered by} Bond et al \citep{bond1991} Press-Schechter\citep{press1974formation}, Sheth-Tormen (ST) \citep{sheth2002excursion}, and the following works \citep{Maggiore:2009rv, Maggiore:2009rw, nikakhtar2018, driver2022, gu2023non}. For simplicity in notation, we omit the subscript $h$ and the redshift dependence in conditional luminosity function.
 In this article, we use the ST mass distribution in EST. 

{{There have been several observational, as well as semianalytical results that show that the conditional luminosity function of central galaxies $	\Phi _c (L|M)$ can be described by a log-normal function \citep{mo2010galaxy}. The log-normal function has a dispersion about $0.15~ \rm{dex}$ \citep{yang2008galaxy}. This dispersion is small enough that we can ignore it in this stage of work, at least for the purpose of modeling the global properties of the luminosity function and its environmental dependence. So, we can write the conditional luminosity function of central galaxies as a delta-function:
\begin{equation}
	\Phi _c (L|M)= \delta_{D}(L-L_c(M)),
	\label{eq:8}
\end{equation} 
where $L_c$ is the luminosity of the central galaxy, which is a function of the host DM halo mass}}.
Extending this conditional luminosity function to all types of galaxies requires us to assume that all the galaxies are considered centrals. This assumption can be efficient given the fact that recent simulation shows the fraction of satellite galaxies inside voids is negligible \citep{melanie2020,yetli2022}. If we make this assumption, we can rewrite equation \ref{eq:7} as 
\begin{equation}
	\Phi(L) = \beta \left[ n(M(L))\frac{dM(L)}{dL}\right]|_{L=L_c},
	\label{eq:9}
\end{equation}
 $\beta$ is the normalization factor that can be extracted using observation or simulation data inserted in equation \ref{eq:7}. We define $\beta$ to capture the environmental dependence of the luminosity function. To proceed further, we need the relation between the mass of the DM halo and the luminosity of the galaxy. The methods, like halo abundance matching, give this relationship \citep{ yang2008galaxy, yang2018elucid}. 
In this formalism, we need to make two basic assumptions \citep{vale2004linking, mo2010galaxy}
(a) Each halo or subhalo hosts only one galaxy.
(b) The luminosity of a galaxy has a monotonic relation with halo mass.
Using these two arguments, we can conclude that the number of galaxies with luminosities higher than $L$ equals the number of halos with masses more than $M$ \citep{vale2004linking}.

\begin{equation}
	\int_L^\infty \Phi(L)dL = \int_M^\infty \left[n_{\rm{h}}(M)+n_{\rm{sh}}(M)\right]dM,
	\label{eq:10}
\end{equation}
where $n_{\rm{h}}(M)$ is number density of main DM halos, and $n_{\rm{sh}}(M)$ is the number density of DM subhalos. Since we are studying the void region, we will neglect the number of subhalos $n_{\rm{sh}}$ at a high mass range of DM halos. 

\begin{table}[t!]
	\centering
	\begin{tabular}{ |c|c|c|c|c|c| } 
		\hline
		$A$ & $M_L$ & $b$ & $c$ & $d$ & $k$\\ 
		\hline
		$5.7\times 10^9 h^{-2} L_{\odot} $ &  $10^{11} h^{-1}M_{\odot}$ & 4 & 0.57 &  3.72 & 0.23 \\ 
		\hline
	\end{tabular}
	\caption{The best-fit parameters of the fitting function for luminosity-mass relation \citep{vale2004linking}. }
	\label{table:1}
\end{table}
Observation or simulation data can be used to plot the mass-luminosity relation. \cite{vale2004linking} have done this and extracted the following fitting function for the mass-luminosity relation,
\begin{equation}
	L(M)=A{(M/M_L)^b\over{(c+(M/M_L)^{dk}})^{1\over k}},
	\label{eq:11}
\end{equation}
in which A, $M_L$, b, c, d, k are free parameters. Table \ref{table:1} shows the best-fit values of these parameters based on \cite{vale2004linking}'s {paper. They have used several observational surveys to obtain the data points for the luminosity of individual galaxies and used these data to further fit the free parameters of equation} \ref{eq:11}. Using this formula and also considering the ST halo mass function in equation \ref{eq:1}, we can calculate the galaxy luminosity function. 
On the other hand, to test this hypothesis and also to find the value of the normalization factor $\beta$, we need a set of data on the galaxy luminosity function.

\section{Results: Environmental and redshift dependence of galaxy luminosity function}\label{Sec3}
To assess equation \ref{eq:9}, {and to also find the value of $\beta$ we use 2dFGR data set as listed in table} \ref{table:2}. {The description of 2dFGRS data set is given in the next subsection. By fitting the analytical formula to this data, we find the best-fit $\beta$ in different regions, as mentioned in table} \ref{table:2}.  As it is shown in Table \ref{table:2}, we can deduce that $\beta$ has a direct dependence on the environment.

\begin{table}[t!]
    \centering
    \begin{tabular}{ |c|c|c| } 
         \hline
         Region Type & $\delta_v$ & Best-Fit $\beta$ \\ 
         \hline
         sparse voids & -0.90 & $0.000406^{+0.000083}_{-0.000100}$\\ 
         \hline
         populous voids & -0.75 & $0.000561^{+0.000050}_{-0.000054}$\\ 
         \hline
         underdense regions & -0.43 & $0.001361^{+0.000083}_{-0.000094}$\\ 
         \hline
         general halo & NA & $0.004450^{+0.000110}_{-0.000092}$\\ 
         \hline
    \end{tabular}
    \caption{Best-Fit values of parameter $\beta$ based on 2dFGRS data with $95\%$ confidence level uncertainty. $\delta_{v}$ is the nonlinear critical density contrast. For convenience, we have named the two void regions, populous and sparse voids, according to their density contrast.}
    \label{table:2}
\end{table}

\begin{table}[t!]
	\centering
	\begin{tabular}{ |c|c|c|c| } 
		\hline
		parameter & $\zeta_1$ & $\zeta_2$ & $\zeta_3$\\
		\hline
		Value & $-0.661\pm 0.032$ &  $-19.547\pm 0.074$ & $-0.081\pm 0.010$ \\
		\hline
		parameter & $\zeta_4$ & $\zeta_5$& \\ 
		\hline
		Value & $0.0149\pm 0.0012$ & $-3.61\pm 0.24$&\\ 
		\hline
	\end{tabular}
	\caption{The best-fit parameter values of the fitting function Equation \ref{eq:14} using 2dFGRS data. The uncertainties are written based on the $68.3\%$ confidence level.}
	\label{table:3}
\end{table}


\subsection{Environmental  dependence of galaxy luminosity function} \label{Sec3.1}
The data set used in this section is extracted from 2dFGRS \citep{croton20052df}. The limiting apparent magnitude of the survey in the $b_j$ band is $m_{b_j}<19.5$. Furthermore, the redshift range of the survey is $z<0.2$.
It should be mentioned that several theoretical models have been proposed to formulate the environmental dependence of the galaxy luminosity function. For example, \cite{mcnaught2014galaxy} has proposed the following model, which is {a standard Schechter function, but one in which the parameters are allowed to vary as a function of density.} They claimed that it can best explain the luminosity function data. 
\begin{eqnarray} 
    \phi\left(M\right)&=&0.4\ln10\times\phi_\ast\times{10}^{0.4\left(M_\ast-M\right)\left(\alpha+1\right)} \label{eq:12} \\ &\times& \exp(-{10}^{0.4\left(M_\ast-M\right)}), \nonumber
\end{eqnarray}
where
\begin{eqnarray}
	&&\phi_* = \gamma_1(1+\delta_m)+\gamma_2, \nonumber \\
	&&M_*=\gamma_3( \log_{10} (1+\delta_m))+\gamma_4,  \label{eq:13}\\
	&&\alpha = \gamma_5, \nonumber
\end{eqnarray}
in which $\gamma_1$ to $\gamma_5$ are free parameters. The density contrast used in the aforementioned equation is in the Eulerian coordinate as mentioned in the corresponding paper. Similar to \cite{mcnaught2014galaxy}, to propose an analytical formulation, we started with the Schechter function \ref{eq:12} and argued that the free parameters should depend on the density contrast of the environment. However, our model only works in the void region. The Schechter parameters of our model are best described by
\begin{figure}[t!]
	\centering
	\includegraphics[width=\columnwidth]{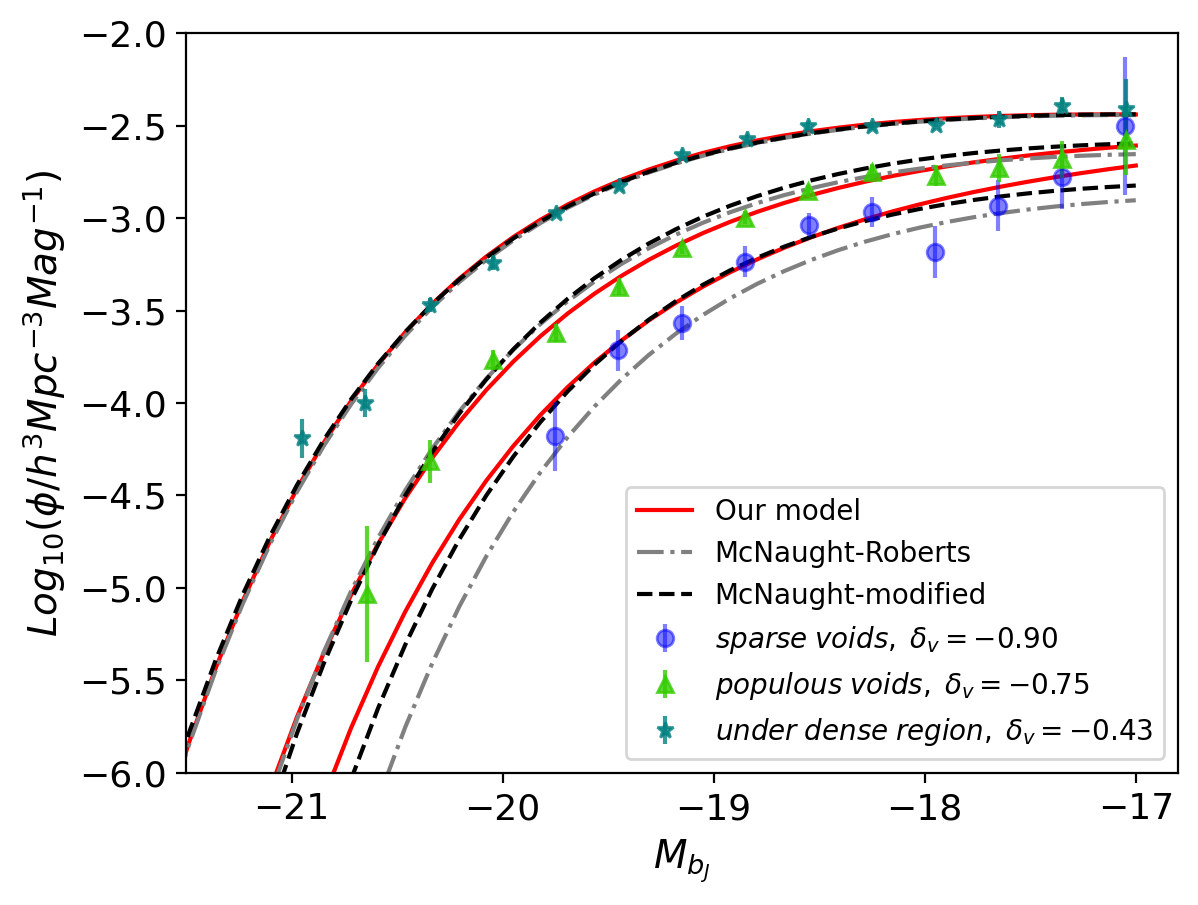}
	\caption{{Galaxy luminosity function as a function of magnitude for three different environments labeled by their density contrast. All of the environments are in the under dense region so we can relate the galaxy luminosity function to the critical density of the host void. The nonlinear (Eulerian) value of the density contrast is mentioned in the figure}. The data points are from 2dfGRS \citep{croton20052df}.}
	\label{fig:1}
\end{figure}
\begin{figure}[t!]
	\centering
	\includegraphics[width=\columnwidth]{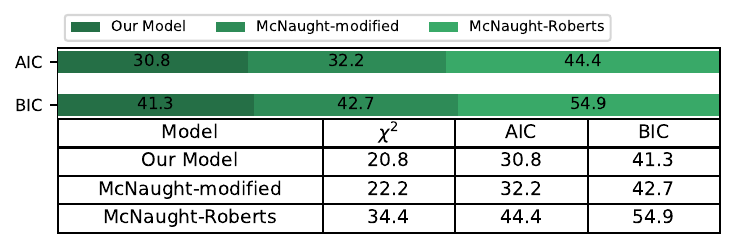}
	\caption{{AIC and BIC test results comparing our environmental dependence model of the galaxy luminosity function, McNaught-Roberts-modified, and McNaught-Roberts models using 2dFGRS data and ADAM optimizer \cite{kingma2014adam}. The results show an apparent difference in the compatibility of the models to the data.}}
	\label{fig:2}
\end{figure}

\begin{figure}[t!]
	\centering
	\includegraphics[width=\columnwidth]{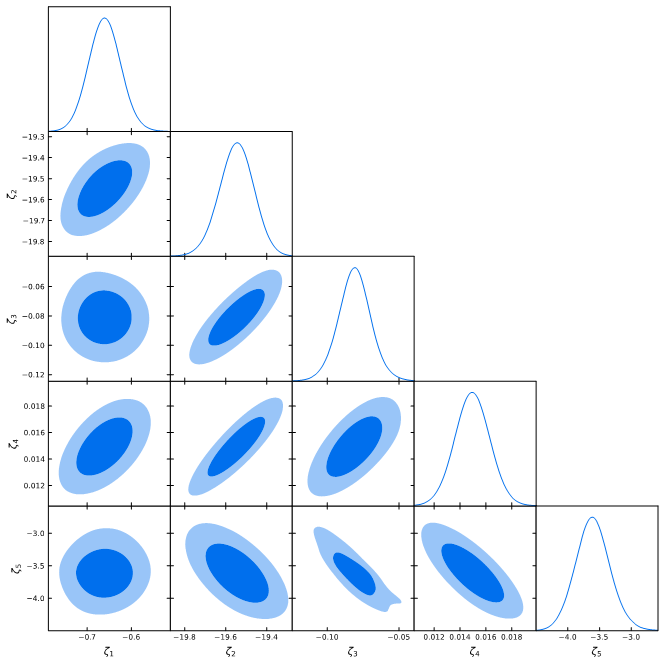}
	\caption{The confidence level of free parameters of Schechter function defined in equation \ref{eq:14}. Dark and light blue are $68\%$ and $95\%$ confidence regions.}
	\label{fig:MCMC}
\end{figure}

\begin{eqnarray}
	&&\phi_* = \zeta_1(1+\mathcal{D})^{\zeta_2}, 	\nonumber\\
	&&M_*=\zeta_3(1+\mathcal{D})^{\zeta_4}, \label{eq:14} \\
	&&\alpha = \zeta_5(1+\mathcal{D}), 	\nonumber
\end{eqnarray}
in which $\mathcal{D}\equiv \frac{\left|\delta_v\right|}{\delta_c+\left|\delta_v\right|}$, where $\delta_c$ is the critical density for halo formation and $\delta_v$ is the threshold of the formation of void (as mentioned in Section \ref{Sec2.1}). Additionally, $\zeta_1$ to $\zeta_5$ are free parameters and their best-fit values (calculated using 2dFGRS data) are mentioned in table \ref{table:3} along with their $68.3\%$ confidence level. The $\chi ^2$ test shows an improvement in comparison with the model previously proposed by \cite{mcnaught2014galaxy}. It should be noted that this comparison is made by using the same set of data (2dFGRS). Figure \ref{fig:1} {shows the 2dFGRS data on the luminosity function as a function of $b_j$-band magnitude, and the lines are the result of fitting our model (equation} \ref{eq:14}) {to this data. This data is presented in three different density contrast thresholds as extracted by the work of} \cite{croton20052df}, which have negative values, and thus we used them to test our model specifically in the void region. Hence, the density contrast dependence of our model (equation \ref{eq:14}) has been assessed to be a more suitable option in the under dense region than McNaught-Roberts et al. (2014) model (equation\ref{eq:13}). {{Since the number of free parameters in both models is the same, the AIC and BIC tests, as shown in figure \ref{fig:2}, also prioritize the new model for the density-dependence of the luminosity function. We should note that \cite{mcnaught2014galaxy}'s proposal is not constrained to the void region. In other words, they have proposed a model based on the density contrast of any environment, regardless of whether it is a void or not. On the other hand, our model is exclusively for the void region. The fact that we have included the environment dependence through the parameter $\mathcal{D}$, could be the reason behind this superiority, as it considers the distance between the two barriers, one related to the halo (in which the galaxy resides) and one related to the larger region that hosts this halo. In fact, the effect of this new parameterization of the density contrast ($\mathcal{D}$), shows a remarkable improvement even when used in the old} \cite{mcnaught2014galaxy}'s model (equation \ref{eq:13}) in comparison to using the $\delta_m$ parameter. Keeping the same form in equation \ref{eq:13} and replacing $\delta_m$ with $\mathcal{D}$ reduces the $\chi^2$ by $35\%$. This version of the model is labeled as "McNaught-modified" in Figure} \ref{fig:2}. \\
In figure \ref{fig:MCMC}, we plot the confidence level contours of the best-fit of the free parameters of our model. In the next subsection, we will study the redshift dependence of luminosity function. 
\subsection{Redshift dependence of luminosity function}
\label{Sec3.2}
Voids are amongst the most frequent types of environments in the low-redshift Universe. That motivated us to analyze the dependence of galaxy luminosity function and redshift inside the void region in the low-redshift limit. Once more, to check whether equation \ref{eq:9} is a suitable approach to link the galaxy luminosity function to the halo mass function, we used a data set extracted from the galaxy and mass assembly (GAMA) survey \citep{gama2015galaxy}. This survey consists of $\sim3 \times 10^5$ galaxies down to an apparent magnitude of $m_r < 19.8$ in the $r$ band. The allowed redshift range for the galaxy templates used in the survey was up to $z\simeq 0.9$. The luminosity functions for a redshift range of $0.04<z<0.26$ are extracted from \cite{gama2014galaxy}. Furthermore, \cite{mcnaught2014galaxy} have shown that the redshift dependence of the galaxy luminosity function is most evident in the underdense region in the low-redshift regime. The redshift dependence of the galaxy luminosity function has been previously analyzed in several studies \citep{bouwens2021new}. It has been preferred to include the redshift dependence of the galaxy luminosity function through the free parameters of the Schechter function. By fitting the luminosity function of galaxies of different redshifts with observational data, one can extract different models to formulate this dependence. But if we want to consider a more theory-based analysis, we go back to equation \ref{eq:7}. As the equation implies, the redshift dependence could enter the equation through the halo mass function. Once again, we limit our study to the void region. Thus, the halo mass function should also represent the number density of halos in these regions. For this purpose, we use an analytical model inspired by equation \ref{eq:5}. Using this equation, along with the halo mass function, we can argue that the redshift dependence of the halo mass function will be of the form:
\begin{equation}
	\Phi(M,z)\propto \frac{\exp(-\eta /{D^2(z)})}{D(z)},
	\label{eq:15}
\end{equation}
in which $D(z)$ is the growth function and $\eta$ is a free parameter. This results in the fact that the redshift dependence only enters the Schechter function through $\phi_*$. This conclusion is in agreement with previous works, which indicated that out of all three parameters of the Schechter function, $\phi_*$ has the most dependence on redshift \citep{loveday2012galaxy,bouwens2021new}. The presented model for the redshift evolution shows deviation from the data trend in higher magnitudes (faint-end). This could be a consequence of quenching on the high star formation end of the main sequence. At higher magnitudes, the faint-end slope parameter of the Schechter function ($\alpha$) could better justify the redshift evolution of the luminosity function.  

\begin{figure}[t!]	
	\centering
	\includegraphics[width=\columnwidth]{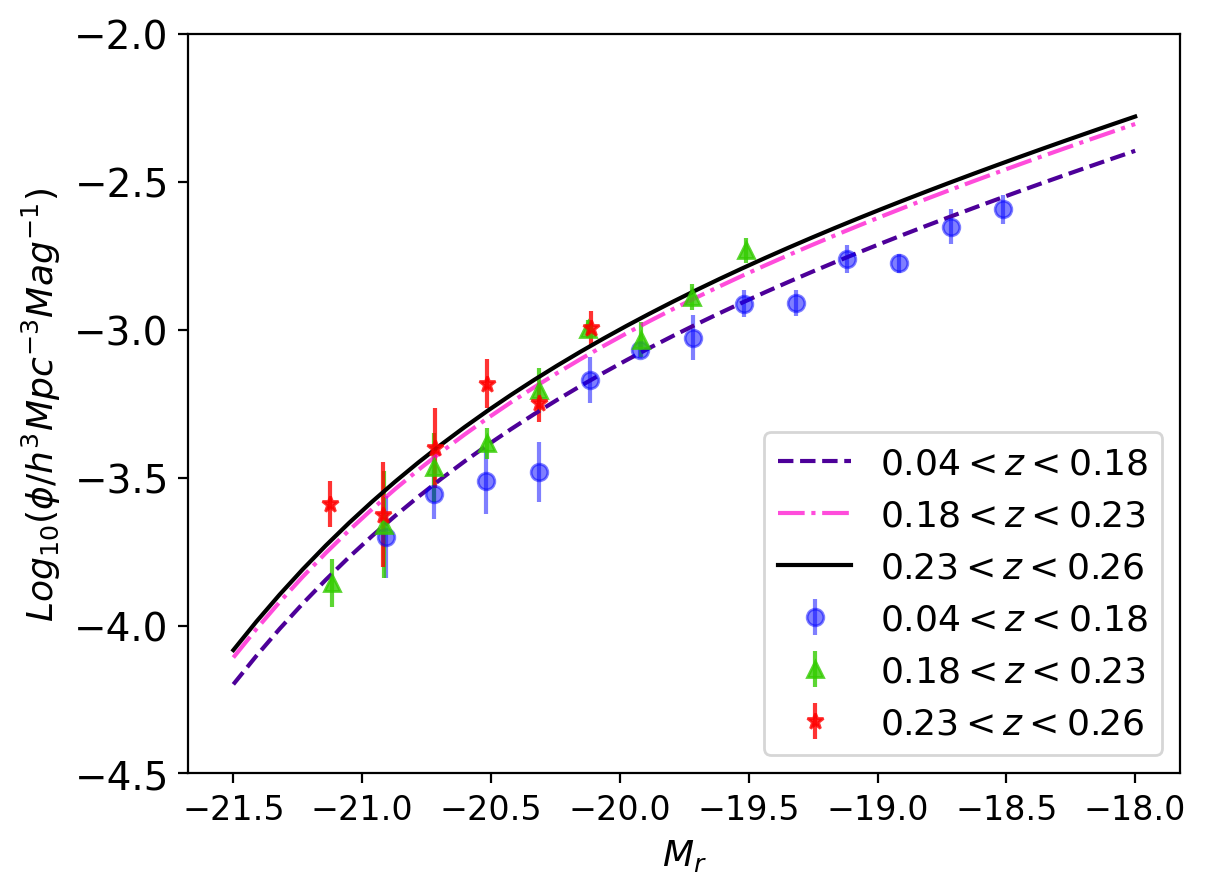}
	\caption{{Galaxy luminosity function as a function of r-band magnitude for three different redshift intervals inside the void region ($-1\leq \delta_v \leq -0.75$). The lines are produced by fitting our model of redshift-dependence} (equation \ref{eq:15}) to GAMA data points. {The data points are taken from \cite{mcnaught2014galaxy} figure 9.}}
	\label{fig:redshift}
\end{figure}

It should be noted that the redshift dependence has entered the halo mass function through the growth function $D(z)$. Using the growth function, we can also analyze the effect of the cosmological model on the galaxy luminosity function. 
We have used a data set from the GAMA survey extracted for the void region \citep{mcnaught2014galaxy} to test this hypothesis. This data set presents three redshift bins: $0.04<z<0.18$, $0.18<z<0.23$, and $0.23<z<0.26$. We have considered the mean redshift value of each bin in equation \ref{eq:13} for fitting. Figure \ref{fig:redshift} shows the proposed theory fitted to the data with the best-fit value $\eta=0.068^{+0.017}_{-0.018}$.


\section{conclusion}
\label{Sec4}
One of the main questions in structure formation is the degree of dependence of galaxies on their host DM halos. This dependence can also shed light on the underlying cosmological model.  In this work, we concentrated on the galaxies that reside in voids. For this task, we investigated the luminosity function of void galaxies. The lack of disturbing processes inside the voids makes it more convenient for us to analyze the properties of galaxy luminosity function.

In this study, we consider the voids as the largest regions in which the density contrast is less than a critical density, as proposed by the shell crossing process. We first, find the coefficient relating the halo mass function to the galaxy luminosity function as described by equation \ref{eq:7}, using 2dFGR and GAMA surveys. We realized that this coefficient is monotonically related to the density contrast of the environment. 

To incorporate the environmental effect in the galaxy luminosity function, we were inspired by EST to include the difference in the linear densities of the enclosing void region and the DM halo inside it (parameterized in $\mathcal{D}$). The result of this approach, which is given in equation \ref{eq:14}, shows an improvement in the $\chi ^2$ value in comparison to the previously proposed models \citep{mcnaught2014galaxy} based on the 2dFGRS dataset \citep{croton20052df}.
It is worth mentioning that the GAMA only provides one density interval in the void region. Hence, it is not feasible to measure the density-dependence of this data set in the underdense environment, and that is why we use 2dfGRS as an alternative.

We also study the redshift dependence of luminosity function. We were again inspired by the EST to include the evolution of the luminosity function through the growth function. According to equation \ref{eq:15}, the redshift dependence which is entered through the halo mass function, is encapsulated in the growth function. Through this, we can use the luminosity function evolution to test the cosmological model. Using the halo mass function to consider the redshift dependence is in agreement with previously proposed papers that suggested a redshift dependence model similar to the {Schechter} function and claimed that its free parameters should be redshift dependent. In \cite{mcnaught2014galaxy}, $\phi _*$ is the parameter that has the strongest redshift dependence. The comparison of this model with data is shown in figure \ref{fig:redshift}. Of course, equation \ref{eq:7} with log-normal distribution of luminosity of central galaxies tells us that the mass and luminosity evolution could be effective in the redshift dependence of the luminosity function. This dependence could be analyzed in future works. Of course, accessing newer and larger sets of data can help us reach a more subtle model and verify the existing ones with more precision. Among the potential surveys and data sets, we can name DESI \citep{wang2024measuringconditionalluminositystellar}, and GALLUMI \citep{sabti2022}. 

\section*{ACKNOWLEDGMENTS}
We thank the anonymous referee, whose insightful comments and detailed suggestions elevated the manuscript to a new level.\\
We would like to thank Farnik Nikakhtar, Arghavan Shafiee, and Ravi K. Sheth for fruitful discussions. \\
SB is partially supported by the Abdus Salam International Centre for Theoretical Physics (ICTP) under the regular associateship scheme.
SB is partially supported by the Sharif University of Technology Office of Vice President for Research under grant No. G4010204.

\bibliography{GalaxyVoid-arXiv-3August2025-published}{}

\begin{thebibliography}{}
\expandafter\ifx\csname natexlab\endcsname\relax\def\natexlab#1{#1}\fi
\providecommand{\url}[1]{\href{#1}{#1}}
\providecommand{\dodoi}[1]{doi:~\href{http://doi.org/#1}{\nolinkurl{#1}}}
\providecommand{\doeprint}[1]{\href{http://ascl.net/#1}{\nolinkurl{http://ascl.net/#1}}}
\providecommand{\doarXiv}[1]{\href{https://arxiv.org/abs/#1}{\nolinkurl{https://arxiv.org/abs/#1}}}

\bibitem[{Bond {et~al.}(1991)Bond, Cole, Efstathiou, \& Kaiser}]{bond1991}
Bond, J., Cole, S., Efstathiou, G., \& Kaiser, N. 1991, ApJ, Part 1 (ISSN
  0004-637X), vol. 379, Oct. 1, 1991, p. 440-460., 379, 440

\bibitem[{Bouwens {et~al.}(2022)Bouwens, Illingworth, Ellis, Oesch, \&
  Stefanon}]{bouwens2022z}
Bouwens, R., Illingworth, G., Ellis, R.~S., Oesch, P., \& Stefanon, M. 2022,
  ApJ, 940, 55

\bibitem[{Bouwens {et~al.}(2021)Bouwens, Oesch, Stefanon, Illingworth,
  Labb{\'e}, Reddy, Atek, Montes, Naidu, Nanayakkara,
  {et~al.}}]{bouwens2021new}
Bouwens, R., Oesch, P., Stefanon, M., {et~al.} 2021, The Astronomical Journal,
  162, 47

\bibitem[{Croton {et~al.}(2005)Croton, Farrar, Norberg, Colless, Peacock,
  Baldry, Baugh, Bland-Hawthorn, Bridges, Cannon, {et~al.}}]{croton20052df}
Croton, D.~J., Farrar, G.~R., Norberg, P., {et~al.} 2005, Monthly Notices of
  the Royal Astronomical Society, 356, 1155

\bibitem[{Curtis {et~al.}(2024)Curtis, McDonough, \& Brainerd}]{curtis2024}
Curtis, O., McDonough, B., \& Brainerd, T.~G. 2024, ApJ, 962, 58

\bibitem[{Driver {et~al.}(2022)Driver, Robotham, Obreschkow, Peacock, Baldry,
  Bellstedt, Bland-Hawthorn, Brough, Cluver, Holwerda, {et~al.}}]{driver2022}
Driver, S.~P., Robotham, A.~S., Obreschkow, D., {et~al.} 2022, Monthly Notices
  of the Royal Astronomical Society, 515, 2138

\bibitem[{Fontanot {et~al.}(2020)Fontanot, De~Lucia, Hirschmann, Xie, Monaco,
  Menci, Fiore, Feruglio, Cristiani, \& Shankar}]{fontanot2020}
Fontanot, F., De~Lucia, G., Hirschmann, M., {et~al.} 2020, Monthly Notices of
  the Royal Astronomical Society, 496, 3943

\bibitem[{Ganeshaiah~Veena {et~al.}(2021)Ganeshaiah~Veena, Cautun, van~de
  Weygaert, Tempel, \& Frenk}]{balletiii2021}
Ganeshaiah~Veena, P., Cautun, M., van~de Weygaert, R., Tempel, E., \& Frenk,
  C.~S. 2021, Monthly Notices of the Royal Astronomical Society, 503, 2280

\bibitem[{Green(2022)}]{Green_2022}
Green, A. 2022, {SciPost} Physics Lecture Notes,
  \dodoi{10.21468/scipostphyslectnotes.37}

\bibitem[{Gu {et~al.}(2023)Gu, Dor, van Waerbeke, Asgari, Mead, Tr{\"o}ster, \&
  Yan}]{gu2023non}
Gu, S., Dor, M.-A., van Waerbeke, L., {et~al.} 2023

\bibitem[{Habouzit {et~al.}(2020)Habouzit, Pisani, Goulding, Dubois,
  Somerville, \& Greene}]{melanie2020}
Habouzit, M., Pisani, A., Goulding, A., {et~al.} 2020, Monthly Notices of the
  Royal Astronomical Society, 493, 899, \dodoi{10.1093/mnras/staa219}

\bibitem[{Hossen {et~al.}(2022)Hossen, Ema, Bolejko, \&
  Lewis}]{hossen2022ringing}
Hossen, M.~R., Ema, S.~A., Bolejko, K., \& Lewis, G.~F. 2022, Monthly Notices
  of the Royal Astronomical Society, 513, 5575

\bibitem[{Kameli \& Baghram(2022)}]{Kameli:2020kao}
Kameli, H., \& Baghram, S. 2022, MNRAS, 511, 1601,
  \dodoi{10.1093/mnras/stac129}

\bibitem[{Kameli \& Baghram(2025)}]{Kameli:2025qyq}
---. 2025.
\newblock \doarXiv{2505.02475}

\bibitem[{Kingma(2014)}]{kingma2014adam}
Kingma, D.~P. 2014, arXiv preprint arXiv:1412.6980

\bibitem[{Libeskind {et~al.}(2018)Libeskind, Van De~Weygaert, Cautun, Falck,
  Tempel, Abel, Alpaslan, Arag{\'o}n-Calvo, Forero-Romero, Gonzalez,
  {et~al.}}]{libeskind2018tracing}
Libeskind, N.~I., Van De~Weygaert, R., Cautun, M., {et~al.} 2018, Monthly
  Notices of the Royal Astronomical Society, 473, 1195

\bibitem[{Liske {et~al.}(2015)Liske, Baldry, Driver, Tuffs, Alpaslan, Andrae,
  Brough, Cluver, Grootes, Gunawardhana, {et~al.}}]{gama2015galaxy}
Liske, J., Baldry, I.~K., Driver, S.~P., {et~al.} 2015, Monthly Notices of the
  Royal Astronomical Society, 452, 2087

\bibitem[{Loveday {et~al.}(2012)Loveday, Norberg, Baldry, Driver, Hopkins,
  Peacock, Bamford, Liske, Bland-Hawthorn, Brough,
  {et~al.}}]{loveday2012galaxy}
Loveday, J., Norberg, P., Baldry, I.~K., {et~al.} 2012, Monthly Notices of the
  Royal Astronomical Society, 420, 1239

\bibitem[{Maggiore \& Riotto(2010{\natexlab{a}})}]{Maggiore:2009rv}
Maggiore, M., \& Riotto, A. 2010{\natexlab{a}}, ApJ, 711, 907,
  \dodoi{10.1088/0004-637X/711/2/907}

\bibitem[{Maggiore \& Riotto(2010{\natexlab{b}})}]{Maggiore:2009rw}
---. 2010{\natexlab{b}}, ApJ, 717, 515, \dodoi{10.1088/0004-637X/717/1/515}

\bibitem[{McNaught-Roberts {et~al.}(2014{\natexlab{a}})McNaught-Roberts,
  Norberg, Baugh, Lacey, Loveday, Peacock, Baldry, Bland-Hawthorn, Brough,
  Driver, {et~al.}}]{mcnaught2014galaxy}
McNaught-Roberts, T., Norberg, P., Baugh, C., {et~al.} 2014{\natexlab{a}},
  Monthly Notices of the Royal Astronomical Society, 445, 2125

\bibitem[{McNaught-Roberts {et~al.}(2014{\natexlab{b}})McNaught-Roberts,
  Norberg, Baugh, Lacey, Loveday, Peacock, Baldry, Bland-Hawthorn, Brough,
  Driver, {et~al.}}]{gama2014galaxy}
---. 2014{\natexlab{b}}, Monthly Notices of the Royal Astronomical Society,
  445, 2125

\bibitem[{Mo {et~al.}(2010)Mo, Van~den Bosch, \& White}]{mo2010galaxy}
Mo, H., Van~den Bosch, F., \& White, S. 2010, Galaxy Formation and Evolution
  (Cambridge University Press)

\bibitem[{Mo \& White(1996)}]{mo1996}
Mo, H., \& White, S.~D. 1996, Monthly Notices of the Royal Astronomical
  Society, 282, 347

\bibitem[{Mo {et~al.}(2004)Mo, Yang, Bosch, \& Jing}]{mo2004dependence}
Mo, H., Yang, X., Bosch, F.~C., \& Jing, Y. 2004, Monthly Notices of the Royal
  Astronomical Society, 349, 205

\bibitem[{Morganti(2017)}]{morganti2017}
Morganti, R. 2017, Frontiers in Astronomy and Space Sciences, 4, 42

\bibitem[{Nikakhtar {et~al.}(2018)Nikakhtar, Ayromlou, Baghram, Rahvar,
  Rahimi~Tabar, \& Sheth}]{nikakhtar2018}
Nikakhtar, F., Ayromlou, M., Baghram, S., {et~al.} 2018, Monthly Notices of the
  Royal Astronomical Society, 478, 5296

\bibitem[{Nikakhtar \& Baghram(2017)}]{Nikakhtar:2016bju}
Nikakhtar, F., \& Baghram, S. 2017, PhRvD, 96, 043524,
  \dodoi{10.1103/PhysRevD.96.043524}

\bibitem[{Parkavousi {et~al.}(2023)Parkavousi, Kameli, \&
  Baghram}]{parkavousi2023voids}
Parkavousi, L., Kameli, H., \& Baghram, S. 2023, Monthly Notices of the Royal
  Astronomical Society, 526, 1495

\bibitem[{Press \& Schechter(1974)}]{press1974formation}
Press, W.~H., \& Schechter, P. 1974, ApJ, Vol. 187, pp. 425-438 (1974), 187,
  425

\bibitem[{Roos(1981)}]{roos1981galaxy}
Roos, N. 1981, Astronomy and Astrophysics, vol. 104, no. 2, Dec. 1981, p.
  218-228., 104, 218

\bibitem[{Rosas-Guevara {et~al.}(2022)Rosas-Guevara, Tissera, Lagos, Paillas,
  \& Padilla}]{yetli2022}
Rosas-Guevara, Y., Tissera, P., Lagos, C. d.~P., Paillas, E., \& Padilla, N.
  2022, Monthly Notices of the Royal Astronomical Society, 517, 712,
  \dodoi{10.1093/mnras/stac2583}

\bibitem[{Sabti {et~al.}(2022)Sabti, Munoz, \& Blas}]{sabti2022}
Sabti, N., Munoz, J.~B., \& Blas, D. 2022, Physical Review D, 105, 043518

\bibitem[{Schechter(1976)}]{schechter1976analytic}
Schechter, P. 1976, Astrophysical Journal, Vol. 203, p. 297-306, 203, 297

\bibitem[{Scholz-D{\'{\i} }az {et~al.}(2022)Scholz-D{\'{\i} }az,
  Mart{\'{\i}}n-Navarro, \& Falc{\'{o}}n-Barroso}]{Scholz_D_az_2022}
Scholz-D{\'{\i} }az, L., Mart{\'{\i}}n-Navarro, I., \& Falc{\'{o}}n-Barroso, J.
  2022, Monthly Notices of the Royal Astronomical Society, 511, 4900,
  \dodoi{10.1093/mnras/stac361}

\bibitem[{Sharma {et~al.}(2022)Sharma, Page, \& Breeveld}]{sharma2022}
Sharma, M., Page, M., \& Breeveld, A. 2022, Monthly Notices of the Royal
  Astronomical Society, 511, 4882

\bibitem[{Sheth {et~al.}(2001)Sheth, Mo, \& Tormen}]{Sheth:1999su}
Sheth, R.~K., Mo, H.~J., \& Tormen, G. 2001, MNRAS, 323, 1,
  \dodoi{10.1046/j.1365-8711.2001.04006.x}

\bibitem[{Sheth \& Tormen(2002)}]{sheth2002excursion}
Sheth, R.~K., \& Tormen, G. 2002, Monthly Notices of the Royal Astronomical
  Society, 329, 61

\bibitem[{Sheth \& Van De~Weygaert(2004)}]{sheth2004hierarchy}
Sheth, R.~K., \& Van De~Weygaert, R. 2004, MNRAS, 350, 517

\bibitem[{Tavasoli {et~al.}(2015)Tavasoli, Rahmani, Khosroshahi, Vasei, \&
  Lehnert}]{tavasoli2015}
Tavasoli, S., Rahmani, H., Khosroshahi, H.~G., Vasei, K., \& Lehnert, M.~D.
  2015, The Astrophysical Journal Letters, 803, L13

\bibitem[{Vale \& Ostriker(2004)}]{vale2004linking}
Vale, A., \& Ostriker, J. 2004, Monthly Notices of the Royal Astronomical
  Society, 353, 189

\bibitem[{Wang {et~al.}(2024)Wang, Yang, Gu, Xu, Xu, Wang, Katsianis, Han, He,
  Zheng, Li, Wang, Hong, Wang, Tan, Zou, Lange, Hahn, Behroozi, Aguilar, Ahlen,
  Brooks, Claybaugh, Cole, de~la Macorra, Dey, Doel, Forero-Romero, Honscheid,
  Kehoe, Kisner, Lambert, Manera, Meisner, Miquel, Moustakas, Nie, Poppett,
  Rezaie, Rossi, Sanchez, Schubnell, Tarlé, Weaver, \&
  Zhou}]{wang2024measuringconditionalluminositystellar}
Wang, Y., Yang, X., Gu, Y., {et~al.} 2024, Measuring the conditional luminosity
  and stellar mass functions of galaxies by combining the DESI LS DR9, SV3 and
  Y1 data.
\newblock \doarXiv{2312.17459}

\bibitem[{Wechsler \& Tinker(2018)}]{Wechsler_2018}
Wechsler, R.~H., \& Tinker, J.~L. 2018, Annual Review of Astronomy and
  Astrophysics, 56, 435, \dodoi{10.1146/annurev-astro-081817-051756}

\bibitem[{White \& Rees(1978)}]{white1978core}
White, S.~D., \& Rees, M.~J. 1978, Monthly Notices of the Royal Astronomical
  Society, 183, 341

\bibitem[{Yang {et~al.}(2008)Yang, Mo, \& van~den Bosch}]{yang2008galaxy}
Yang, X., Mo, H., \& van~den Bosch, F.~C. 2008, ApJ, 676, 248

\bibitem[{Yang {et~al.}(2009)Yang, Mo, \& Van~den Bosch}]{yang2009galaxy}
Yang, X., Mo, H., \& Van~den Bosch, F.~C. 2009, ApJ, 695, 900

\bibitem[{Yang {et~al.}(2018)Yang, Zhang, Wang, Liu, Lu, Li, Shi, Jing, Mo,
  van~den Bosch, {et~al.}}]{yang2018elucid}
Yang, X., Zhang, Y., Wang, H., {et~al.} 2018, ApJ, 860, 30

\bibitem[{Zentner(2007)}]{Zentner:2006vw}
Zentner, A.~R. 2007, IJMPD, 16, 763, \dodoi{10.1142/S0218271807010511}

\bibitem[{Zhang {et~al.}(2022)Zhang, Yang, \& Guo}]{Zhang_2022}
Zhang, Y., Yang, X., \& Guo, H. 2022, Monthly Notices of the Royal Astronomical
  Society, 517, 3579, \dodoi{10.1093/mnras/stac2934}

\end{thebibliography}
\bibliographystyle{aasjournal}

\end{document}